\documentclass[iop]{emulateapj}
\usepackage{amsmath}
\usepackage{multirow}
\usepackage{graphicx}
\usepackage{color}
\newcommand{\kms}{km s$^{-1}$}
\newcommand{\tCO}{$^{13}$CO~2-1}
\newcommand{\sgn}{\mathop{\mathrm{sgn}}}

\slugcomment{Accepted to ApJ Letters: 5 March, 2014}

\shorttitle{Spiral arms in the disk of HD~142527}
\shortauthors{Christiaens et al.}

\begin{document}

\title{Spiral arms in the disk of HD~142527 from CO emission lines with ALMA}

%% Use \author, \affil, and the \and command to format
%% author and affiliation information.
%% Note that \email has replaced the old \authoremail command
%% from AASTeX v4.0. You can use \email to mark an email address
%% anywhere in the paper, not just in the front matter.
%% As in the title, use \\ to force line breaks.

\author{V. Christiaens\altaffilmark{1} , S. Casassus\altaffilmark{1}, S. Perez\altaffilmark{1}, G. van der Plas\altaffilmark{1} \& F. M\'enard\altaffilmark{2}}
\email{valchris@das.uchile.cl}

%\author{C. D. Biemesderfer\altaffilmark{4,5}}
%\affil{National Optical Astronomy Observatories, Tucson, AZ 85719}
%\email{aastex-help@aas.org}
%
%\and
%
%\author{R. J. Hanisch\altaffilmark{5}}
%\affil{Space Telescope Science Institute, Baltimore, MD 21218}

%% Notice that each of these authors has alternate affiliations, which
%% are identified by the \altaffilmark after each name.  Specify alternate
%% affiliation information with \altaffiltext, with one command per each
%% affiliation.

\altaffiltext{1}{Departamento de Astronom\'ia, Universidad de Chile, Casilla 36-D, Santiago, Chile.}
\altaffiltext{2}{UMI-FCA, CNRS/INSU, France (UMI 3386), and Dept. de Astronom\'{\i}a, Universidad de Chile, Santiago, Chile.}
%\altaffiltext{2}{Society of Fellows, Harvard University.}

\begin{abstract}

In view of both the size of its gap and the previously reported
asymmetries and near-infrared spiral arms, the transition disk of the
Herbig Fe star HD~142527 constitutes a remarkable case study.  This
paper focuses on the morphology of the outer disk through ALMA
observations of $^{12}$CO J=2-1, $^{12}$CO J=3-2 and $^{13}$CO J=2-1.
Both $^{12}$CO J=2-1 and $^{12}$CO J=3-2 show spiral features of
different sizes.  The innermost spiral arm (S1) is a radio counterpart
of the first near-infrared spiral observed by \citet{Fukagawa2006},
but it is shifted radially outward.  However, the most conspicuous CO
spiral arm (S2) lies at the outskirts of the disk and had not been
detected before.  It corresponds to a cold density structure, with both 
brightness and excitation temperatures
of order 13$\pm2$ K and conspicuous in the $^{12}$CO J=2-1 peak-intensity
map, but faint in $^{12}$CO J=3-2.  There is also a faint counterarm
(S3), point-symmetrical of S2 with respect to the star. These three
spirals are modelled separately with two different formulae that
approximate the loci of density maxima in acoustic waves due to
embedded planets.  S1 could be fit relatively well with these
formulae, compared to S2 and S3. Alternative scenarios such as
gravitational instability or external tidal interaction are discussed. 
The impact of channelization on spectrally and spatially resolved 
peak intensity maps is also briefly addressed.

\end{abstract}

%\email{\myemail}

%% From the front matter, we move on to the body of the paper.
%% In the first two sections, notice the use of the natbib \citep
%% and \citet commands to identify citations.  The citations are
%% tied to the reference list via symbolic KEYs. The KEY corresponds
%% to the KEY in the \bibitem in the reference list below. We have
%% chosen the first three characters of the first author's name plus
%% the last two numeral of the year of publication as our KEY for
%% each reference.

\section{INTRODUCTION}

%Importance of transition disks

The young circumstellar disks that host a large annular gap or central cavity, 
the so-called \emph{transition disks} (TDs), %are believed to
%be in an intermediate state between protoplanetary and debris disks. They 
could be crucial in the context of planetary formation as their
morphologies may result from dynamical clearing due to planet(s)
\citep{Dodson-Robinson2011, Zhu2012b, Zhu2013a}, rather than dust
grain growth or photo-evaporation alone \citep{Rosotti2013}.

% in particular as their lifetime is a key parameter for constraining
%the two main giant planet formation scenarios: core accretion
%\citep[e.g.][]{Lissauer1993, Pollack1996} and gravitational
%instability \citep[e.g.][]{Boss1997, Boss1998}.  It has been shown
%that gap or cavity opening can be explained in principle by dust
%grain growth and/or photo-evaporation \citep[e.g.][]{Birnstiel2012,
%Owen2012}.  Nevertheless, those two phenomena cannot reproduce the
%largest observed gaps and cavities, so that ,

%that may be of low mass \citep{Zhu2013a}, or one single giant planet located close to the outer edge of the gap/cavity \citep{Zhu2012b}.
%This paradigm is supported by recent discoveries of planetary companion candidates within TDs \citep{Huelamo2011, KrausIreland2012, Arnold2012}. %Isella 2013: suggest az. asymmetry of disk intensity due to planets

% Asymmetries and spirals in young circumstellar disks

Asymmetric features, warps or spirals in TDs can also evidence 
the presence of stellar or planetary companions \citep[e.g.][]{GoldreichTremaine1979}.  
Spiral features have mostly been detected in optical or near-infrared (NIR)
wavelengths \citep[e.g.][]{Grady2001, Clampin2003, Fukagawa2004,
  Muto2012, Grady2013}, and in the disk of AB~Aur at radio wavelength
\citep{Corder2005, Lin2006}, although in
counter-rotation with the disk, thus probably stemming from a late envelope
infall \citep{Tang2012}.

%,  LinPapaloizou1986}. 
%As for warps and asymmetries, the case of
%$\beta$ Pic's edge-on disk is particularly revealing. After the
%detection of several asymmetries \citep{KalasJewitt1995}
%and a warp \citep{Mouillet1997} in the disk, a planetary
%companion was eventually found and confirmed \citep{Lagrange2009,
%  Lagrange2010}.  On the other hand, 

%Specific case of HD142527

The disk around the Herbig Fe star \objectname{HD~142527} constitutes
an outstanding example of a nearly face-on TD \citep[$i$$\sim$28$\pm$3\degr;][]{Perez2014},
showing both a large asymmetric dust depleted gap and a spiral arm
observed in NIR scattered light \citep{Fukagawa2006}. Its mass and age were estimated to respectively
$\sim$1.9-2.2 M$_{\odot}$ and 1-12 Myr old \citep{Fukagawa2006, Verhoeff2011}, for
a distance of 145$\pm$15 pc \citep{DeZeeuw1999, Acke2004}.
\citet{Fujiwara2006} concluded that the disk's rotation axis was
inclined so that the east side was the far side, with a position angle
(PA) of $\sim$-20\degr.  With K-band imaging, \citet{Casassus2012} reported the presence of smaller spiral
structures at the outer edge of the gap, confirmed by
\citet{Rameau2012} and in NIR polarized intensity studies \citep[][the
  latter work identifies two new spirals]{Canovas2013, Avenhaus2014}.

ALMA provided substantial insight into the structure of
HD~142527. \citet{Casassus2013} found gap-crossing
HCO$^+$(4-3) filaments, diffuse CO(3-2) inside the cavity, and
confirmed the horseshoe morphology of the dust continuum reported by
\citet{Ohashi2008}, which they interpreted as a dust trap.
\citet{Fukagawa2013} reported on another ALMA dataset in $^{13}$CO
J=3-2, C$^{18}$O J=3-2 and underlying continua. 
Here we focus on the outskirts of the \objectname{HD 142527} disk % using the ALMA band~6
%and band~7 data presented by \citet{Perez2014} and \citet{Casassus2013} respectively. 
and report on the discovery of several CO spiral arms.

%A new large scale spiral arm, brighter in CO 2-1, is discovered, along with a roughly axisymmetric counterarm.
%A CO counterpart, slightly shifted radially outward, to the scattered light spiral arm of \citet{Fukagawa2006} is also detected. 
%%Attention is drawn on the spurious segmentation of the peak intensity maps.
%The observed features are then further characterized based on two different mathematical formalisms in an attempt to explain them by the presence of a dynamic perturbator.
%Least-square fits result in degenerate values for the parameters of the spirals...
%Other origins of the spiral arms, such as gravitational instability of the disk, are briefly discussed.
%The formation of the large scale spiral arm by gravitational instability (GI) is finally shown to be equally viable.
%through Monte-Carlo based LIME software (REF).

\section{OBSERVATIONS}

\subsection{Datasets}

The observations were obtained with ALMA on June 2012 (Cycle 0), and cover
$^{12}$CO J=2-1 (230.538 GHz, hereafter CO 2-1), $^{12}$CO J=3-2
(345.796 GHz, CO 3-2) and $^{13}$CO J=2-1 (220.399 GHz, \tCO), with 
baselines comprised between 21 (resolution of $\sim$12.8\arcsec) and 379 m ($\sim$0.71\arcsec).
For details on the instrumental setup and calibration we refer to \citet{Perez2014} and \citet{Casassus2013},
respectively for bands~6 (CO 2-1 and \tCO) and 7 (CO 3-2).

Spectral line imaging was achieved with the \verb+CASA+ package.  The
continuum was first subtracted in the visibility domain.  We 
%converted the observed frequencies to the Local Standard of Rest (LSR), and
binned the original spectral channels to 0.2 \kms~bin-width. Next,
visibility data were \textit{clean}ed (Cotton-Schwab
\verb+CLEAN+).  As faint extended structures were seeked,
natural weighting was preferred over Briggs weighting, and masks were drawn 
manually in the dirty maps for each velocity channel. To avoid spurious detections, the masks delimited signal at $\sim$5$\sigma$.
For comparison, an elliptical mask identical in all velocity channels, embracing the whole butterfly pattern, 
yielded similar moment maps.

\subsection{Moment Maps}\label{sec:MomentMaps}

Inspection of the channel maps for the different CO transitions
%(Figure~\ref{fig:bChannelMaps}) 
confirms the LSR systemic radial
velocity of 3.7$\pm$0.2~\kms~\citep{Casassus2013, Perez2014}. Both
CO 2-1 and CO 3-2 (but not \tCO) channels with velocity higher than the systemic
velocity show a decrement in the
specific intensity levels between 4.2-4.8~\kms (\emph{southern} velocities). 
These velocities
correspond to an extended diffuse cloud best detected in single-dish
data \citep[][]{Casassus2013AA...553A..64C}%, and were hence
%omitted for the computation of the integrated intensity maps
. 

\begin{figure*}[htb]
\centering
\includegraphics[width=0.49\textwidth, angle=270]{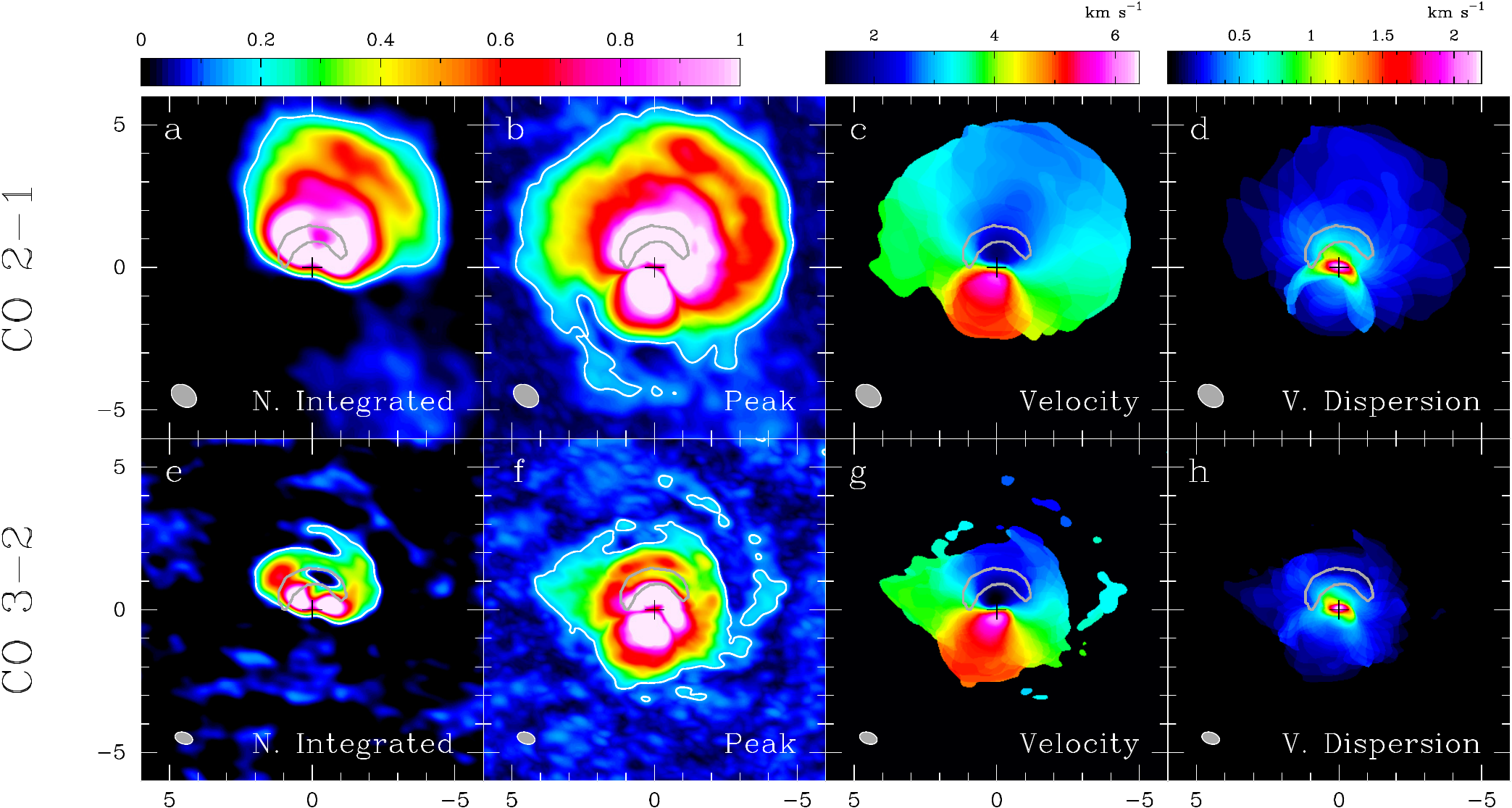}
\caption{Moment maps for CO 2-1 and CO 3-2. Respective beam sizes are 
  0.96$\times$0.72 and 0.63$\times$0.40 arcsec$^2$. North
  is up and east is to the left. $x-$ and $y$ axis indicate angular
  offset from the star along R.A. and Dec.  from the position of the star.
  % at the epoch of observations, i.e. J2000 $15^{\mathrm{h}}56^{\mathrm{m}}41^{\mathrm{s}}.878$ -42\degr19\arcmin23\arcsec.568. 
  A continuum contour at 180 mJy beam$^{-1}$ \citep[half-maximum, from][]{Casassus2013} is overplotted in
  \emph{light gray}.
  {\em Velocity-integrated intensity maps}: \textit{\bf
    a} and \textit{\bf e}. For CO 2-1 and CO 3-2, the 1~$\sigma$ noise levels are respectively 14 and
  17 mJy beam$^{-1}$; and the color scales are both normalized 
  to 20 $\sigma$. %Only northern velocities are included. 
  The 3$\sigma$ contour is shown in white. {\em
    Peak-intensity maps}: \textit{\bf b} and
  \textit{\bf f}. With $I_\mathrm{peak}$ maps, the mean of the noise is non-null; 
  it equals respectively 27 and 42 mJy beam$^{-1}$ for CO 2-1 and CO 3-2.
  The 1~$\sigma$ values are respectively 11 and
  15 mJy beam$^{-1}$; and the color scales are both normalized 
  to 40 $\sigma$.  The relevant contour, mean(noise)+3$\sigma$, is shown in white.
    {\em Velocity centroid}: \textit{\bf c} and \textit{\bf g}. 
    {\em Velocity dispersion}: \textit{\bf d} and \textit{\bf h}.
\label{fig:maps}}
\end{figure*}

For CO~2-1 and CO~3-2 lines, maps of the velocity-integrated intensity ($I_\mathrm{int}$), 
the peak intensity ($I_\mathrm{peak}$), and the velocity 
centroid and dispersion are computed in order to further
characterize the outer disk (Figure~\ref{fig:maps}). The
intensity scale is chosen to emphasize the fainter structure in the
outer disk, at the expense of saturation in the brighter central
regions, discussed in \citet{Casassus2013} and \citet{Perez2014}. We restrict
the velocity range to $2.6 < v_\mathrm{lsr} / \mathrm{km~s}^{-1} <3.6$
(northern velocities) for the computation of the
$I_\mathrm{int}$ maps because of the intervening cloud%, except for \tCO as no decrement in the southern velocities is detected
.

\begin{figure*}[htb]
\centering
\includegraphics[width=0.35\textwidth, angle=270]{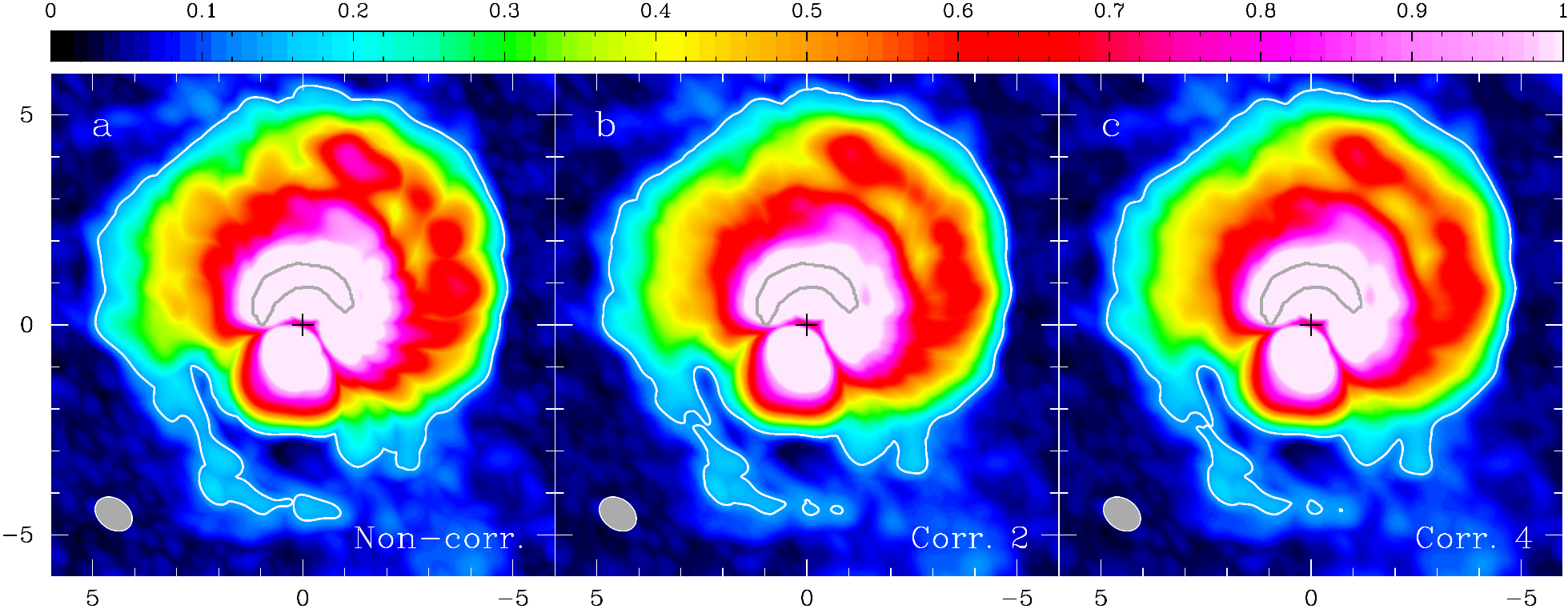} 
\caption{CO 2-1 $I_\mathrm{peak}$ maps: unsmoothed (\textit{\bf a}),
  smoothed using two (\textit{\bf b}) and smoothed using four
  (\textit{\bf c}) shifted datacubes.  Legends follow from
  Figure~\ref{fig:maps}\textit{b}. The rest of the paper implicitly
  shows $I_\mathrm{peak}$ maps smoothed using four shifted
  datacubes. \label{fig:segmentation}}
\end{figure*}

The computation of the $I_\mathrm{peak}$ maps revealed an unknown
issue%that had not been reported in the literature, to our knowledge
. As seen in Figure~\ref{fig:segmentation}a, the signal in the disk appears
segmented for CO~2-1 (this is also the case for CO~3-2).  This might 
be due to the discretization of
the velocity field in channel-averaged visibilities.  To
prove that the segmentation is spurious, we computed two
(resp. four) $I_\mathrm{peak}$ maps with the same velocity bin width
(0.2~km~s$^{-1}$), but shifted by half (resp. a quarter of) the bin
width, and then averaged them.  The fragmentation follows the velocity
channels, and is almost
completely smoothed-out when averaging the different binnings
(Figure~\ref{fig:segmentation}b and c).

%The \emph{smoothed} $I_\mathrm{peak}$ maps were calculated from two
%(resp. four) datacubes starting respectively at -4.0 and -3.9 \kms~
%(resp. -4.00, -3.95, -3.90 and -3.85 \kms).

From the CO 2-1 $I_\mathrm{int}$ map in the
north (Figure~\ref{fig:maps}a), an arc-like structure in the
NW of the image, already hinted in the channel maps, 
is detected at very large distance from the star ($\sim$3.6\arcsec).
Closer-in, at about 2.4\arcsec, the intensities are higher
and more extended to the NNW than to the NNE. In the
$I_\mathrm{peak}$ map (Figure~\ref{fig:maps}b), the closer-in structure
turns out to be roughly shaped as a spiral arm (hereafter S1), whereas
the large scale arc is revealed as a conspicuous tightly wound spiral (S2).  
At an approximately point-symmetric location of
S2 with respect to the star, we also detect a remarkable counterspiral
(S3).

We estimated an upper limit on the flux loss due to interferometer
filtering by aperture photometry in the relevant channels of a LIME \citep{Brinch2010} model 
of HD~142527 \citep{Perez2014}. Cleaning the model, sampled at the uv-coverage of our observation, and
comparing with the original model
yields flux losses $\lesssim$16\%. %(for the systemic velocity
%channel) and 10\% (higher velocity channels).
%This is an
%upper limit to the amount of flux lost since the spirals are
%expected on higher spatial frequencies than a smooth, axisymmetric
%disk.
While this result suggests that filtering-out is minor for S2, the faintness of S3 stems probably from
absorption due to the intervening cloud.

%A fraction, probably large, of the signal coming from S3 is missing, 
%filtered-out by the interferometer, due to the intervening cloud, hence its faintness.  
%\footnote{As we are dealing with $I_\mathrm{peak}$ maps, the mean of their noise is non-null. Their exact values are provided in the caption of Figure~\ref{fig:maps}.}.
% Mean of the spiral in $I_\mathrm{peak}$ map: 219 mJy [Polygon statistics in ds9]
% Median of the spiral in $I_\mathrm{peak}$ map: 230 mJy. 
Bridge-like features in the central regions of the $I_\mathrm{peak}$
maps (Figure~\ref{fig:maps}$b$ and $f$), also seen in the velocity dispersion map
(Figure~\ref{fig:maps}$d$ and $h$), are due to this intervening
cloud%, almost opaque in $^{12}$CO
. The kinematics in CO 2-1 appear to be Keplerian, even under spirals S1
and S2.  The spirals do not have counterparts in velocity dispersion,
i.e. there is no local enhancement of the velocity gradient or
turbulence.

The CO 3-2 $I_\mathrm{int}$ map in the {\it north}
(Figure~\ref{fig:maps}e) displays S1 more evidently, and the
$I_\mathrm{peak}$ map (Figure~\ref{fig:maps}f) also confirms its presence.
Some faint extended patches of signal, just above the
mean(noise)+3$\sigma$ level also follow relatively well S2 detected in
Figure~\ref{fig:maps}b.  %As further detailed in a forthcoming paper, the
%CO 3-2 velocity profile (Figure~\ref{fig:maps}g) is \emph{twisted}
%compared to CO 2-1 (Figure~\ref{fig:maps}c) and
%\tCO~(Figure~\ref{fig:maps}k).

However, the outer disk seems too faint to enable detection of extended structures in \tCO.

\subsection{Spiral arms}

The three spiral arms extend, in projected angular separations but
deprojected physical distances:

\begin{itemize}
\item from $\sim$1.9\arcsec ($\sim$290 AU) at PA$\sim$-110\degr~to $\sim$2.8\arcsec ($\sim$380 AU) at PA$\sim$0\degr~for S1;  % deproj 2.3''-3.0'' , nope rather: 2.0-2.6''
\item from $\sim$3.0\arcsec ($\sim$520 AU) at PA$\sim$-100\degr~to $\sim$4.2\arcsec ($\sim$640 AU) at PA$\sim$0\degr~for S2; % deproj 3.6''-4.4''  ,  OK - OK
\item from $\sim$3.2\arcsec ($\sim$520 AU) at PA$\sim$100\degr~to $\sim$4.4\arcsec ($\sim$670 AU) at PA$\sim$190\degr~for S3. % deproj 3.6''-4.6'',  OK - +0.2
\end{itemize}		
		
They all subtend $\sim$100\degr~in azimuth. S2 and S3 have radial widths
ranging 0.6-1.1\arcsec~(90-160 AU), and are thus globally resolved radially. 
%Considering $i \sim 28$\degr, their physical radial width ranges %0.7-1.25\arcsec~(
%100-180 AU%)
%.

%This wideness can be both
%explained by the beam size effect and the inclination
%$i\sim28$\degr. ****SO THEY SHOULD BE EVEN WIDER?*** ***GIVE SIZE
%RANGE AND SAY IF RESOLVED FOR SURE, OR NOT****

\subsubsection{Physical Diagnostics in the Spiral Arms}\label{sec:DescriptionSpirals}
		
Except for S3, whose signal is absorbed, several physical parameters
describing the spirals may be derived. In $I_\mathrm{peak}$(CO 2-1),
the maximum intensities are $\sim$450 mJy beam$^{-1}$ for S1 (at
PA$\sim$-70\degr, R$\sim$330~AU), and $\sim$280 mJy beam$^{-1}$ for S2 (at
%$\sim$300 mJy beam$^{-1}$ for S2 (at
PA$\sim$-80\degr, R$\sim$530~AU).  In $I_\mathrm{peak}$(CO
3-2), the maximum intensities, roughly at the same locations as for CO 2-1, 
are $\sim$320 mJy beam$^{-1}$ for S1, and $\sim$120 mJy beam$^{-1}$ for S2.

Assuming that $^{12}$CO is optically thick, %the Rayleigh-Jeans formula applied to 
the peak intensities provide brightness
temperatures, which approximate the kinetic temperatures at the
unit-opacity surfaces, of $T_{b,S1}$$\sim$20 K (from both CO 2-1 and CO 3-2 maximum intensities) 
and $T_{b,S2}$$\sim$11-15 K (the upper value stems from CO 2-1 maximum intensity).
Assuming LTE, the excitation temperatures derived from the ratio of CO 2-1 and CO 3-2 
maximum intensities are $T_{ex,S1}$$\sim$22-27~K and $T_{ex,S2}$$\sim$13-15~K.
%(this may stem from an average kinetic temperature lower in the regions where 3-2 originates).

%However, assuming LTE, the
%excitation temperatures derived from the ratio of CO 2-1 and CO 3-2
%maximum intensities are respectively $T_\mathrm{ex,S1}\sim24\pm3$ K and
%$T_\mathrm{ex,S2}\sim14\pm1$ K. We see that
%$T_\mathrm{ex}$ is systematically higher than $T_b$, which indicates
%mild opacities of $\sim 1$.
 %The unit opacity surface for 3-2 would
%ie below 2-1, despite its higher upper-level energy.
%stronger flux-loss in CO 3-2. 
%If so, increasing $I_\mathrm{peak}(\mathrm{CO}~
%3-2)$ so that $T_b(3-2) = T_b(2-1)$, gives an excitation temperature
%of 35-36 K for both spirals. Beam dilution would further increase the observed
%$T_b(\mathrm{CO}~2-1)$, compared to 3-2. 
%An alternative interpretation is that the average kinetic temperature is lower in the
%regions where 3-2 originates. The unit opacity surface for 3-2 would
%lie below 2-1, despite its higher upper-level energy. This may indeed
%be the case since 3-2 is less optically thick than 2-1\footnote{The Einstein A 
%coefficient is $10^{-5.60266}$ for CO 3-2 and $10^{-6.16050}$ for 2-1 (values retrieved 
%from \url{splatalogue.net/}).}.
%In any case, 

CO gas is expected to condensate on dust for T$\lesssim$19-20~K
\citep[e.g.][]{Qi2011,deGregorioMonsalvo2013}. The observation of CO gas at lower
temperature in S2 may suggest that dust grains are depleted, or mainly settled in the
mid-plane \citep[e.g.][]{DullemondDominik2004b}, or that CO desorption
is efficient at the surface
\citep[][]{Hersant2009AA...493L..49H}, at that radial distance. 
%{\bf Unfortunately, we were not able to constrain the dust-to-gas ratio to discriminate among the different possibilities.}
%Unfortunately the available
%continuum observations are not deep enough to quantify the possibility
%of dust depletion at large radii.

%The rms noise in the 230.5~GHz continuum is 1.2~mJy~beam$^{-1}$, which
%for a standard opacity \citep[][]{Beckwith1990AJ.....99..924B} at 15~K
%corresponds to 1.0~M$_\oplus$ of dust per beam. By contrast, the peak
%CO 2-1 emission under S2 corresponds to $10^{-4}~$M$_\oplus$ of CO gas
%per beam, or about 1.0~M$_\oplus$ of gas per beam, so that the limits
%on the gas-to-dust ratio is not constraining.

The gas temperature in the gap is $\sim$41-43~K \citep{Fukagawa2013, Perez2014}.
Using this value at 80~AU, 20-27~K at 330~AU ($T_{b,S1}$-$T_{ex,S1}$) and 
13-15~K at 530~AU ($T_{b,S2}$-$T_{ex,S2}$), the best fit to the law $T
\propto r^{-q}$ yields $q$$\sim$0.5-0.6. This range of values is
steeper than $q$$\sim$0.3 found by \citet{Perez2014} up to $\sim$300~AU.
%, {\bf but closer to what is expected from passive irradiation ($T \propto r^{-0.5}$).}
%, \textcolor{blue}{and are indicative of ??? ***not sure what to write, Simon: "shadowing by puffed-up inner rim", Seba: "shadowing by inner disk", 
%Gerrit: "a bit daring to say it indicates shadowing"***}.

Sound speed and
scale height are defined respectively as $c_s^2 = k_B T_b / (\mu m_p)$
and $H = c_s / \Omega$, where $k_B$ is the Boltzmann constant; $\mu$
is the mean molecular weight of the gas in terms of proton mass $m_p$,
namely $\sim$2.3 for H$_2$ with 10\% of He; and $\Omega$ is the angular frequency at the
physical radius where the temperature is derived% ($\sim$330 AU and $\sim$580 AU for S1 and S2 respectively)
. With $T_{S1}$$\sim$20-27~K and
$T_{S2}$$\sim$13-15~K, and assuming Keplerian rotation, we find scale heights of $H_{S1}$$\sim$38-44 AU and
$H_{S2}$$\sim$66-76 AU, hence a disk aspect ratio $h_S$$\sim$0.12-0.13
at the location of both spirals.
\citet{Verhoeff2011} quoted a scale height of 60~AU at the inner rim of the outer disk ($h$$\sim$0.46).
However, considering a disk with $i$$\sim$28$\pm$3\degr~ \citep{Perez2014} instead of 20\degr~leads to a scale height of 20$\pm$7~AU ($h$$\sim$0.15$\pm$0.05)  %cf. p B67 of 
in the inclination/inner rim scale height relation of \citet{Avenhaus2014}, thus in agreement with our result.
%Interestingly, although the
%outer-disk scale height follows a standard power-law index of \textcolor{blue}{1.25 
%with radius \citep{Verhoeff2011}, the aspect ratio is at least four times
%smaller under S1 and S2 than at the inner rim \citep[0.10 compared 
%to 0.46;][]{Verhoeff2011}}.

We can also obtain lower limits on the mass of each spiral in LTE. The
integrated flux was computed in S1 and S2 from $I_\mathrm{peak}$(CO
2-1) by defining an aperture 1\arcsec~wide in radius, and centered on
the best fit model spiral to \citet{Kim2011} equation (see section \ref{sec:ModellingSpirals}).
We obtain integrated fluxes of 1.41~Jy
and 1.27~Jy for S1 and S2 respectively.  Using these values with
$T_{ex,S1}$ and $T_{ex,S2}$, a velocity dispersion of
$\sim$0.3~\kms~(see Figure~\ref{fig:maps}d), and assuming [H$_2$]/[CO] =
10$^4$, the total gas mass in each spiral is $M_{S1} \gtrsim 1.38
\times 10^{-6} M_{\odot}$ and $M_{S2} \gtrsim 1.23 \times 10^{-6}
M_{\odot}$.
%the LTE CO gas masses are $3.4 \times 10^{21}$ kg and
%$2.9 \times 10^{21}$ kg in S1 and S2 respectively.  

\subsubsection{Modelling of the Spiral Arms}\label{sec:ModellingSpirals}

\begin{figure*}[htb]
\centering
\includegraphics[width=0.7\textwidth]{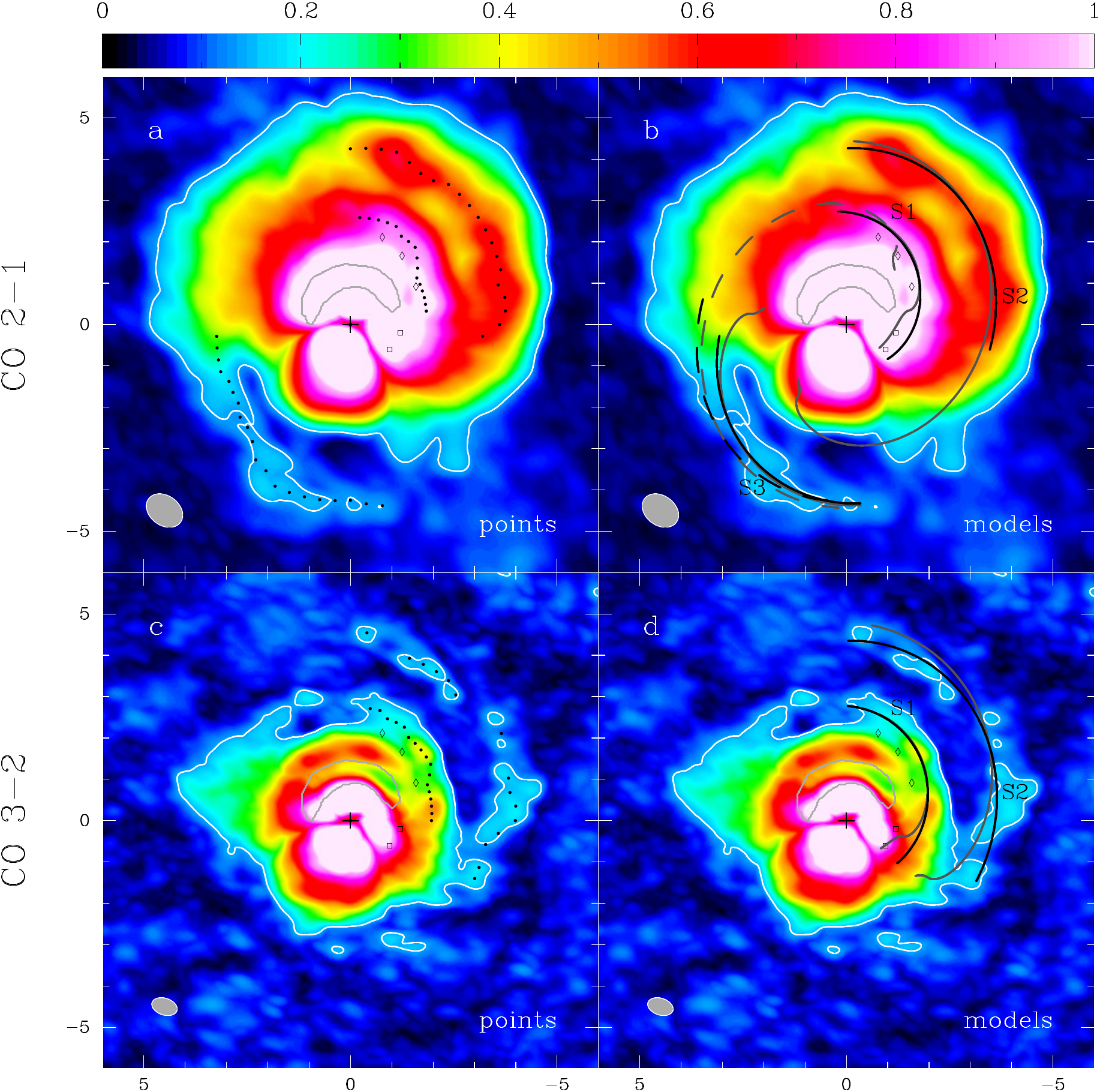}
\caption{Modelling of the spirals observed in the CO 2-1 and CO 3-2
  $I_\mathrm{peak}$ maps. Legends follow from Figure~\ref{fig:maps}$b$
  and $f$. For comparison, we indicate the position of the H-band
  spiral arm from \citet[][{\it diamonds}]{Fukagawa2006}, and the Ks-band spiral root
  from \citet[][{\it squares}]{Casassus2012}.{\bf (a)}: Points tracing the spirals used
  for the modelling. {\bf (b)}: Modelling of the spiral arms as
  in \citet[{\it solid dark gray lines}]{Muto2012} and \citet[{\it
      solid black lines}]{Kim2011}. The {\it dashed dark gray} and
  {\it dashed black} spirals represent the point-symmetric of S2
  models with respect to the star.  {\bf(c)} and {\bf (d)}: Identical
  to (a) and (b) with the CO 3-2 $I_\mathrm{peak}$ map.
\label{fig:bcomparison}}
\end{figure*}

An elliptical grid is defined to match the inclination and PA of the
projected disk, with deprojected bins of 0.125\arcsec~in radius and
5\degr~in azimuth.
The trace of each spiral is identified with radial cuts in
$I_\mathrm{peak}$, using local maxima for S2 and S3, and local inflection 
points for S1 (second-derivative nulls).  We obtain 16, 20 and 20 points tracing
respectively S1, S2, and S3 in the CO 2-1 $I_\mathrm{peak}$ map
(Figure~\ref{fig:bcomparison}a); and 17 and 14 points respectively
tracing S1 and S2 in the CO 3-2 $I_\mathrm{peak}$ map
(Figure~\ref{fig:bcomparison}c).  Uncertainties in the measured
positions of the spirals were set to max\{bin size, quarter of the beam size\}. We
then perform least-squares fits to each spiral independently using two
different parametric formulae, as follows.

%\footnote{A few points were weighted to 2$\sigma$ error bars because
%they deviated strongly from the visual trace of the spirals in
%Figure~\ref{fig:bcomparison}$a$ or $c$.}

The polar coordinates ($\theta$, $r$) of an acoustic wave created by a planet, 
at location ($\theta_c$, $r_c$), on a gaseous disk
can be approximated by \citep[][ and references therein]{Muto2012}:
\begin{align}
\label{Eq:Muto}
\theta(r) &= \theta_c + \frac{\sgn(r-r_c)}{h_c} \nonumber \\ &\times
\Bigg\{ \left( \frac{r}{r_c} \right)^{1+\beta} \bigg[
  \frac{1}{1+\beta} - \frac{1}{1-\alpha +\beta} \left( \frac{r}{r_c}
  \right)^{-\alpha} \bigg] \nonumber \\ & - \left( \frac{1}{1+\beta} -
\frac{1}{1-\alpha +\beta} \right) \Bigg\}
\end{align}
where $h_c$ is the disk aspect ratio
at radius $r_c$.  It is assumed that $\Omega$ and $c_s$ can be
expressed as power laws: $\Omega \propto r^{-\alpha}$ and $c_s \propto
r^{-\beta}$$\propto r^{-q/2}$.  In total, there are thus 5 parameters to each fit
($r_c$, $\theta_c$, $h_c$, $\alpha$ and $\beta$). Note that a
plus sign was used in Eq. \eqref{Eq:Muto} after $\theta_c$ instead of the
minus sign given in \citet{Muto2012} since the disk of \objectname{HD~142527} 
is rotating clockwise.

First fits with equation~\ref{Eq:Muto} were computed by fixing the values
of $\alpha$ to 1.5 (Keplerian rotation) and $\beta$ to 0.25
(see $q$ in section \ref{sec:DescriptionSpirals}).  We found
plausible values of $h_c$ (between 0.01 and 1.0) with the points of CO 2-1
$I_\mathrm{peak}$ map for S1 ($h_c$ = 0.15)
and S3 ($h_c$ = 0.27), but not for S2, and for neither spiral with the points of CO 3-2
$I_\mathrm{peak}$ map.  In order to further reduce the
parameter space, a second set of fits was run fixing as well the value of
$h_c$ to 0.15. The best fit models are shown in \emph{solid dark gray}
lines in Figure~\ref{fig:bcomparison}b and d; the inflection point in the spiral curves  
represents the best fit location of the planet. The values of the best fit
$r_c$, $\theta_c$, and $\chi^2$ are given in Table
\ref{tab:parameters}.

Sweeping $\beta$ from 0.15 \citep{Perez2014} to 0.30 ($q=0.6$)
does not significantly change $\chi^2$ (relative to
the spread of values for different spirals) and the value of the other
parameters.  On the contrary,
the best fit parameters depends strongly on $h_c$; a difference of a few hundredths induces significantly different results, confirming the degeneracy already noted by \citet{Muto2012} and \citet{Boccaletti2013}.
%Attempts with $h_c$ set to 0.12 (found in section \ref{sec:DescriptionSpirals}) yield significantly different values of $r_c$ and $\theta_c$ but only slightly larger $\chi^2$ for the different spirals, confirming the degeneracy already noted by \citet{Muto2012} and \citet{Boccaletti2013}.
  %In view of the parameters degeneracy, it is not surprising that the best fit parameters of CO 2-1 and CO 3-2 points do not match \textcolor{blue}{but they do corroborate within $\sim$2$\sigma$}. 
Since $\chi^2$ depends on the estimated error bars, which
are not independently determined, we use $\chi^2$ to compare different
models rather than assess their statistical significance.  The best
fit spiral appears to be S1, while S2 and S3 provide $\chi^2$ values a
factor of several times larger.

\begin{table*}
\begin{center}
\caption{Values of the best fit parameters with equations \eqref {Eq:Muto} and \eqref{Eq:Kim}.}
\label{tab:parameters}
\begin{tabular}{clccc}
\\
\hline
\hline
\multicolumn{2}{c}{Equation \eqref{Eq:Muto}} & S1 & S2 & S3 \\
\hline
\multirow{3}{*}{CO 2-1} & $r_c$ [arcsec] & 1.71 $\pm$0.04 & 2.13 $\pm$0.24 & 2.62 $\pm$0.08\\
& $\theta_c$ [deg] & 283 $\pm$5 & 146 $\pm$64 & 83 $\pm$10 \\
& $\chi^2$ & 2.38 & 18.0 & 4.66 \\
\hline
\multirow{3}{*}{CO 3-2} & $r_c$ [arcsec] & 1.62 $\pm$0.04 & 2.69 $\pm$0.38 & / \\
& $\theta_c$ [deg] & 255 $\pm$7 & 235 $\pm$45 & / \\
& $\chi^2$ & 2.05 & 36.0 & / \\
\\
\hline
\multicolumn{2}{c}{Equation \eqref{Eq:Kim}} & S1 & S2 & S3 \\
\hline
\multirow{3}{*}{CO 2-1} & a [arcsec rad$^{-1}$] & 0.58 $\pm$0.05 & 0.20 $\pm$0.05 & 0.65 $\pm$0.06\\
& b [arcsec] & -0.86 $\pm$0.29 & 3.05 $\pm$0.29 & 2.34 $\pm$0.16 \\
& $\chi^2$ & 0.16 & 0.30 & 0.40 \\
\hline
\multirow{3}{*}{CO 3-2} & a [arcsec rad$^{-1}$] & 0.45 $\pm$0.03 & 0.25 $\pm$0.10 & / \\
& b [arcsec] & -0.02 $\pm$0.17 & 2.88 $\pm$0.52 & / \\
& $\chi^2$ & 0.18 & 2.94 & / \\
\\
\end{tabular}
\end{center}
$^a$ For Eq. \eqref{Eq:Muto}, we fixed $\alpha = 1.5$ (Keplerian
rotation), $\beta = 0.25$ (from $q$ found in section \ref{sec:DescriptionSpirals}) and $h_c = 0.15$ (best fit value when $h_c$ is set free for S1
in the CO 2-1 $I_\mathrm{peak}$ map).

\end{table*}

Another formula suggested by the theoretical work of \citet{Kim2011} 
approximates the shape of the spiral wake created by an embedded planet on a
circular orbit to an Archimedean spiral:%, each tracing the edges of
%the spiral overdensity:
\begin{equation}
\label{Eq:Kim}
r(\theta) = a \theta + b
\end{equation}
where $a$ corresponds to $r_p / M_p$ with $r_p$ the launching radius
of the planet and $M_p$ its Mach number, and $b$ is a constant.
%depending on whether the inner or outer edge is considered. In our
%case, a single $b$ is chosen to model the center of the overdensity,
%and not the edges. 
Parameters for the best fit of each spiral are
given in Table \ref{tab:parameters}.  %They are less degenerate and lead to better reduced $\chi^2$ than Eq.~\eqref{Eq:Muto}.  
With a $\chi^2$ value at least twice smaller than the two other spirals, 
it is again S1 that is best fit.
Equations \ref{Eq:Muto} and \ref{Eq:Kim} are fundamentally different - $\theta (r)$
vs $r(\theta)$, so that $\chi^2$ obtained with each equation should not be mutually compared.
%Comparing
%$\chi^2$ values between the different spirals, we find that it is
%again S1 that is best fit, with a $\chi^2$ value a factor of at least
%twice smaller than the two other spirals.
%However, the two formulae are fundamentally different - $\theta (r)$
%vs $r(\theta)$ - and the position of the planet is more degenerate
%azimuthally than radially, so that $\chi^2$ values obtained with each
%equation should not be mutually compared.
%For S1, $a$ ranges between 0.45 and 0.58, obtained respectively from
%CO 3-2 and CO 2-1 $I_\mathrm{peak}$ maps points (Table
%\ref{tab:parameters}).  For S2, we find lower values of $a = 0.20
%-0.25$, which corresponds to a more tightly wound spiral.  
%For S3, $a\sim0.65$ reflects a larger pitch angle.  The best fit spirals are shown
%in \emph{solid black} lines in Figure~\ref{fig:bcomparison}b and d.
%The reduced number of parameters in Eq.~\ref{Eq:Kim} allows to reach some conclusions in the Archimedes formalism. 

The similar slopes $a$ of S3 and S1 may 
suggest that S3 is prolongating S1.
Conversely, models for S2 and S3 do not share the same slope, even though 
the point-symmetric of S2 with respect
to the star roughly coincides with S1 and S3 (\emph{dashed dark gray} curve in Figure \ref{fig:bcomparison}b).

\subsubsection{Comparison with NIR Spiral Arms}\label{sec:ComparisonDetections}

Although S2 and S3 had not been detected before, a spiral arm similar to S1
had already been reported in NIR H and K bands by \citet{Fukagawa2006}.  Also in NIR,
closer-in observations in Ks-band \citep{Casassus2012}, L'-band
\citep{Rameau2012}, H- and K-bands polarimetric intensities
\citep{Canovas2013, Avenhaus2014}, all led independently to
the detection of spiral structures closer to the star, originating at
the outer edge of the gap.  In Figure~\ref{fig:bcomparison} we register
the contour crests from the spiral of \citet[][their Figure~2]{Fukagawa2006}, as well as the
most prominent spiral feature from \citet[][spiral \#2 in their Figure~2]{Casassus2012}.
%We notice a striking correspondence between the H-band spiral arm of
%\citet{Fukagawa2006}, its Ks-band root from \citet{Casassus2012}, and the inner rim of S1.  
Our S1 model extends the radio spiral to its NIR
root, suggesting that the H-band spiral arm, its Ks-band root, and the
inner rim of S1 are part of the same structure. %\textcolor{blue}{The NIR root of the spiral 

\section{DISCUSSION: ORIGIN OF THE SPIRAL ARMS}\label{sec:discussion} % ~700 words

Several scenarios have been proposed for the occurrence of spirals in
TDs \citep[see e.g.][for a summary]{Boccaletti2013}. HD~142527,
with its well-defined tightly coiled radio spirals in Keplerian rotation,
provides an interesting case study suggesting a different origin than the CO spirals in the disk of AB~Aur.

%\subsection{Planet or low-mass stellar companion}

%As shown in section \ref{sec:ModellingSpirals}, 
S1 could be fit with
equations~\eqref{Eq:Muto} and \eqref{Eq:Kim} assuming an embedded
companion (section \ref{sec:ModellingSpirals}), although its location on the model spiral
could not be precisely constrained. %\textcolor{blue}{
An object has been 
discovered by \citet{Biller2012} and re-detected in H$\alpha$ by \citet{CloseFollette2014}
%Close et al.~(2014; accepted in ApJL) 
at $\sim$12~AU. Extending
S1 inwards does not allow us to confirm a possible relationship. Nevertheless,
%Besides, 
the clearing of a gap as
large as 140~AU should involve several planets
\citep[e.g.][]{Dodson-Robinson2011}. For S2 and S3, both the
\emph{relatively} poor least-square fit with equations \eqref{Eq:Muto}
and \eqref{Eq:Kim} and their very large scales argue in favor of a
different cause.%}

% While several surveys in
%HD~142527 yielded negative results, a candidate low-mass stellar
%companion has been suggested by \citep{Biller2012} and confirmed by
%Close et al.  (2014; accepted in ApJL), at $\sim$12~AU. The extension
%of S1 inwards does not allow us to confirm a possible relationship
%with their object.  
%

%\subsection{Past encounter or fly-bys}

%%\citet{Toomre1972} showed that the fly-by of a galaxy close to another
%%could create a two-armed temporary spiral pattern by tidal
%%interaction, further enhanced by disk self-gravity.  

%A two-armed spiral pattern such as S2 and S3 could be caused by a
%fly-by \citep[e.g.][]{Toomre1972}. 
Simulations in the context of
protoplanetary disks have shown that two-armed spirals, such as S2 and S3, could be
induced by stellar encounters \citep[e.g.][]{Larwood2001,Quillen2005}.
Similarly, simulations of \citet{Augereau2004} and \citet{Quillen2005}
reproduce the large scale ($\sim$325 AU) tightly wound spiral observed
in the disk of \objectname{HD~141569~A}, by tidal induction from one
of its M-dwarf companions (external to the disk).  Tidally induced
spiral structures are transient and vanish within a few orbits, which
implies either a very recent encounter or a bound companion external
to the disk periodically exciting spirals.  For HD~142527 no such
object has been detected.

%Follow-up is needed to
%determine whether the candidate at $\sim$12~AU could be
%on both an eccentric and inclined orbit enabling the
%excitation of one or some of the observed spirals
%as such.

%Type I and II migration of planets do also create spirals, in addition to reduce the surface density and creating a gap by tidal torque Voir Armitage and Rice 2005! + Gap clearing by a planet in Rice et al 2003.

%\subsection{Gravitational instability}
		
Gravitational instability (G.I.) is an alternative scenario able to
create a grand-design two-armed spiral structure \citep[see
  e.g.][]{Boss1998}. The stability of a disk is characterized by
Toomre's parameter, $Q$ \citep{Toomre1964}.  \citet{Fukagawa2013}
computed an upper limit of 1-2 for $Q$ in the dust continuum
horseshoe, so that depending on the gas-to-dust ratio, fragmentation
of either the gas or the dust or both components due to G.I. may
occur.  Approximating $Q \approx \frac{M_{\star}}{M_d} h_S$
\citep{Gammie2001}, with the mass of HD 142527 $M_{\star}$ set to 2
$M_{\odot}$, the mass of the disk $M_d$ set to 0.1 $M_{\odot}$ \citep{Verhoeff2011} and the
aspect ratio at both spirals $h_S$$\sim$0.1 (section
\ref{sec:DescriptionSpirals}), we find $Q$$\sim$2.0 similarly to
\citet{Fukagawa2013}. This suggests that the outer disk in general,
not just the horse-shoe, is stable but close to the instability
regime.

	\acknowledgments

This paper makes use of the following ALMA data:\\ {\tt
ADS/JAO.ALMA\#2011.0.00465.S}. ALMA is a partnership of ESO, NSF,
NINS, NRC, NSC, ASIAA. The Joint ALMA Observatory is operated by
ESO, AUI/NRAO and NAOJ. Financial support was provided by Milenium
Nucleus P10-022-F (Chilean Ministry of Economy)  and by FONDECYT grant 1130949.
The authors acknowledge the referee for his comments.

%The authors acknowledge the referee for his helpful comments.
%This research was funded by Millenium Science Initiative, Chilean
%Ministry of Economy, Nucleus P10-022-F.
%This paper makes use of the following ALMA data:
%ADS/JAO.ALMA\#2011.0.00465.S. ALMA is a
%partnership of ESO (representing its member states), NSF (USA) and NINS
%(Japan), together with NRC (Canada) and NSC and ASIAA (Taiwan), in
%cooperation with the Republic of Chile. The Joint ALMA Observatory is
%operated by ESO, AUI/NRAO and NAOJ.

%{\it Facilities:} \facility{ALMA}.

%\input{Biblio}
% 42/50 references (v. 15/01/2014)
%\bibliographystyle{apj} % ie, style for natbib! Other example: apacite, apalike

\end{document}